\documentclass[twocolumn,english,prl]{revtex4}
\usepackage[colorlinks=true,urlcolor=blue,citecolor=blue,linkcolor=blue,breaklinks=true]{hyperref} 
\usepackage[T1]{fontenc}
\usepackage[latin9]{inputenc}
\usepackage{amssymb}
\usepackage{graphicx}
\usepackage{amsmath,color}
\usepackage{mathrsfs}
\usepackage{float}
\usepackage{indentfirst}
\usepackage{braket}

\def\journal #1, #2, #3, 1#4#5#6{{\sl #1~}{\bf #2}, #3 (1#4#5#6) }

\def\eqa{\begin{eqnarray}}
\def\eea{\end{eqnarray}}
\newcommand{\eq}{\begin{equation}}
\newcommand{\ee}{\end{equation}}

\newcommand{\Eq}[1]{Eq.~(\ref{#1})}

\newcommand{\rua}{\color{red} \uparrow}
\newcommand{\bda}{\color{blue} \downarrow}

\makeatother
\usepackage{babel}

\begin{document}
 
\title{Discovering Phase Transitions with Unsupervised Learning}

\author{Lei Wang}

\affiliation{Beijing National Lab for Condensed Matter Physics and Institute
of Physics, Chinese Academy of Sciences, Beijing 100190, China}

\begin{abstract}
Unsupervised learning is a discipline of machine learning which aims at discovering patterns in big data sets or classifying the data into several categories without being trained explicitly. We show that unsupervised learning techniques can be readily used to identify phases and phases transitions of many body systems. Starting with raw spin configurations of a prototypical Ising model, we use principal component analysis to extract relevant low dimensional representations the original data and use clustering analysis to identify distinct phases in the feature space. This approach successfully finds out physical concepts such as order parameter and structure factor to be indicators of the phase transition. We discuss future prospects of discovering more complex phases and phase transitions using unsupervised learning techniques. 
\end{abstract}
\maketitle

Classifying phases of matter and identifying phase transitions between them is one of the central topics of condensed matter physics research. Despite an astronomical number of constituting particles, it often suffices to represent states of a many-body system with only a few variables. For example, a conventional approach in condensed matter physics is to identify order parameters via symmetry consideration or analyzing low energy collective degree of freedoms and use them to label phases of matter~\cite{Anderson:1984td}. 

However, it is harder to identify phases and phase transitions in this way in an increasing number of new states of matter, where the order parameter may only be defined in an elusive nonlocal way~\cite{Wen:2004vr}. These new developments call for new ways of identifying appropriate indicators of phase transitions.  

 
To meet this challenge, we use machine learning techniques to extract information of phases and phase transitions directly from many-body configurations. In fact, application of machine learning techniques to condensed matter physics is a burgeoning field~\cite{Curtarolo:2003ema, Ovchinnikov:2009hh, Hautier:2010fu, Snyder:2012da, Saad:2012gf, LeDell:2012gm, Rupp:2012kx, Arsenault:2014em, Pilania:2015dl, Li:2015eb, Carrasquilla:2016wu}\footnote{We also note application of physics ideas such as phase transition~\cite{Anonymous:0Ung1RBG}, renormalization group~\cite{Anonymous:IsmowpLi}, tensor networks~\cite{Anonymous:9rpSZ_B7} and quantum computation~\cite{Lloyd:2013us} to machine learning.}. For example, regression approaches are used to predict crystal structures~\cite{Curtarolo:2003ema}, to approximate density functionals~\cite{Snyder:2012da}, and to solve quantum impurity problems~\cite{Arsenault:2014em}; artificial neural networks are trained to classify phases of classical statistical models~\cite{Carrasquilla:2016wu}. However, most of those applications use \emph{supervised} learning techniques (regression and classification), where a learner needs to be trained with the previously solved data set (input/output pairs) before it can be used to make predictions. 

On the other hand, in the \emph{unsupervised} learning, there is no such explicit training phase. The learner should by itself find out interesting patterns in the input data. Typical unsupervised learning tasks include cluster analysis and feature extraction. Cluster analysis divides the input data into several groups based on certain measures of similarities. Feature extraction finds a low-dimensional representation of the dataset while still preserving essential characteristics of the original data. Unsupervised learning methods have broad applications in data compression, visualization, online advertising and recommender system, etc. They are often being used as a preprocessor of supervised learning to simplify the training procedure. In many cases, unsupervised learning also lead to better human interpretations of complex datasets. 


In this paper, we explore the application of unsupervised learning in many-body physics with a focus on phase transitions. The advantage of unsupervised learning is that one assumes neither the presence of the phase transition nor the precise location of the critical point. Dimension reduction techniques can extract salient features such as order parameter and structure factor from the raw configuration data. Clustering analysis can then divide the data into several groups in the low-dimensional feature space, representing different phases. Our studies show that unsupervised learning techniques have great potentials of addressing the big data challenge in the many-body physics and making scientific discoveries. 

As an example, we consider the prototypical classical Ising model
\begin{equation}
H = -J \sum_{\braket{i,j}} \sigma_{i}\sigma_{j}, 
\label{eq:Ising}
\end{equation}
where the spins take two values $\sigma_{i}=\{-1,+1\}$. 
We consider the model (\ref{eq:Ising}) on a square lattice with periodic boundary conditions and set $J=1$ as the energy unit.  
The system undergoes a phase transition at temperature $T/J={2}/{\ln(1+\sqrt{2})}\approx 2.269$~\cite{Onsager:1944fq}. A discrete $Z_{2}$ spin inversion symmetry  is broken in the ferromagnetic phase below $T_{c}$ and is restored in the disordered phase at temperatures above $T_{c}$. 

We generate $100$ uncorrelated spin configuration samples using Monte Carlo simulation~\cite{Wolff:1989iy} at temperatures $T/J=1.6,1.7,\ldots,2.9$ each and
collect them into an $M\times N$ matrix
\begin{equation}
X = \left(\begin{array}{ccccccc}\rua & \bda & \rua & \ldots & \rua & \rua & \rua \\ &   &   & \vdots &  & & \\ \bda & \rua & \bda & \ldots & \rua & \bda & \rua  \end{array}\right)_{M\times N}, 
\label{eq:X}
\end{equation}
where $M=1400$ is the total number of samples, and $N$ is the number of lattice sites. The up and down arrows in the matrix denote $\sigma_{i}=\pm 1$. Such a matrix is the \emph{only} data we feed to the unsupervised learning algorithm. 

Our goal is to discover possible phase transition of the model (\ref{eq:Ising}) without assuming its existence. This is different from the supervised learning task, where exact knowledge of $T_{c}$ was used to train a learner~\cite{Carrasquilla:2016wu}. Moreover, the following analysis does not assume any prior knowledge about the lattice geometry and the Hamiltonian. We are going to use the unsupervised learning approach to extract salient features in the data and then use this information to cluster the samples into distinct phases. Knowledge about the temperature of each sample and the critical temperature $T_{c}$ of the Ising model is used to verify the clustering. 

Interpreting each row of $X$ as a coordinate of an $N$-dimensional space, the $M$ data points form a cloud centered around the origin of a hypercube~\footnote{Each column of $X$ sums up to zero since on average each site has zero magnetization.}. Discovering a phase transition amounts to find a hypersurface which divides the data points into several groups, each representing a phase. The task is akin to the standard unsupervised learning technique: cluster analysis~\cite{Anonymous:eeFaU9Du}, where numerous algorithms are available, and they group the data based on different criteria. 

\begin{figure}
  \centering
  \includegraphics[width=\columnwidth]{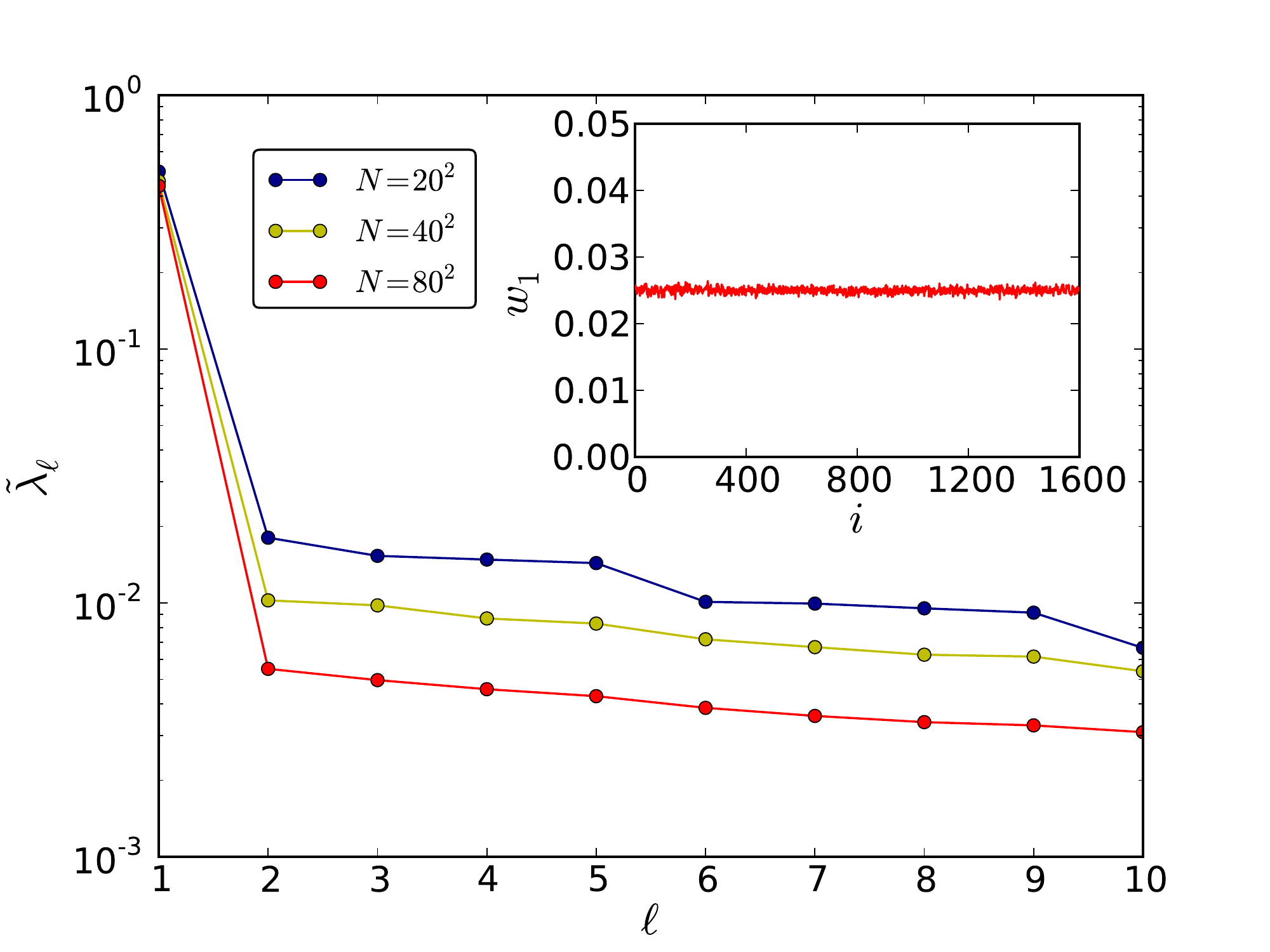}
  \caption{The first few explained variance ratios obtained from the raw Ising configurations. The inset shows the weights of the first principal component on an $N=40^{2}$ square lattice.}
  \label{fig:components}
\end{figure}

However, direct applying clustering algorithms to the Ising configurations may not be very enlightening. The reasons are twofold. First, even if one manages to separate the data into several groups, clusters in high dimensional space may not directly offer useful physical insights. Second, many clustering algorithms rely on a good measure of similarity between the data points. Its definition is, however, ambiguous without supplying of domain knowledge such as the distance between two spin configurations. 

On the other hand, the raw spin configuration is a highly redundant description of the system's state because there are correlations among the spins. Moreover, as the temperature varies, there is an overall tendency in the raw spin configurations, such as lowering the total magnetization. In the following, we will try to first identify some crucial features in the raw data. They provide an effective low dimensional representation of the original data. And in terms of these features, the meaning of 
the distance between configurations becomes more transparent.   
The separation of phases is also often clearly visible and comprehensible by the human in the reduced space spanned by these features. Therefore, feature extraction does not only simplifies the subsequent clustering analysis but also provides effective means of visualizing and offering physical insights. We denote the crucial features extracted by the unsupervised learning as \emph{indicators} of the phase transition. In general, they do not necessarily need to be the same as the conventional order parameters defined in condensed matter physics. This unsupervised learning approach nevertheless provides an alternative view of phases and phase transitions. 






Principal component analysis (PCA)~\cite{Pearson:1901gs} is a widely used feature extraction technique. The principal components are mutually orthogonal directions along which the variances of the data decrease monotonically. PCA finds the principal components through a linearly transformation of the original coordinates $Y = X W$. When applied to the Ising configurations in \Eq{eq:X}, PCA finds the most significant variations of the data changing with the temperature. We interpret them as relevant features in the data and use them as indicators of the phase transition if there is any. 

\begin{figure}
  \centering
  \includegraphics[width=\columnwidth]{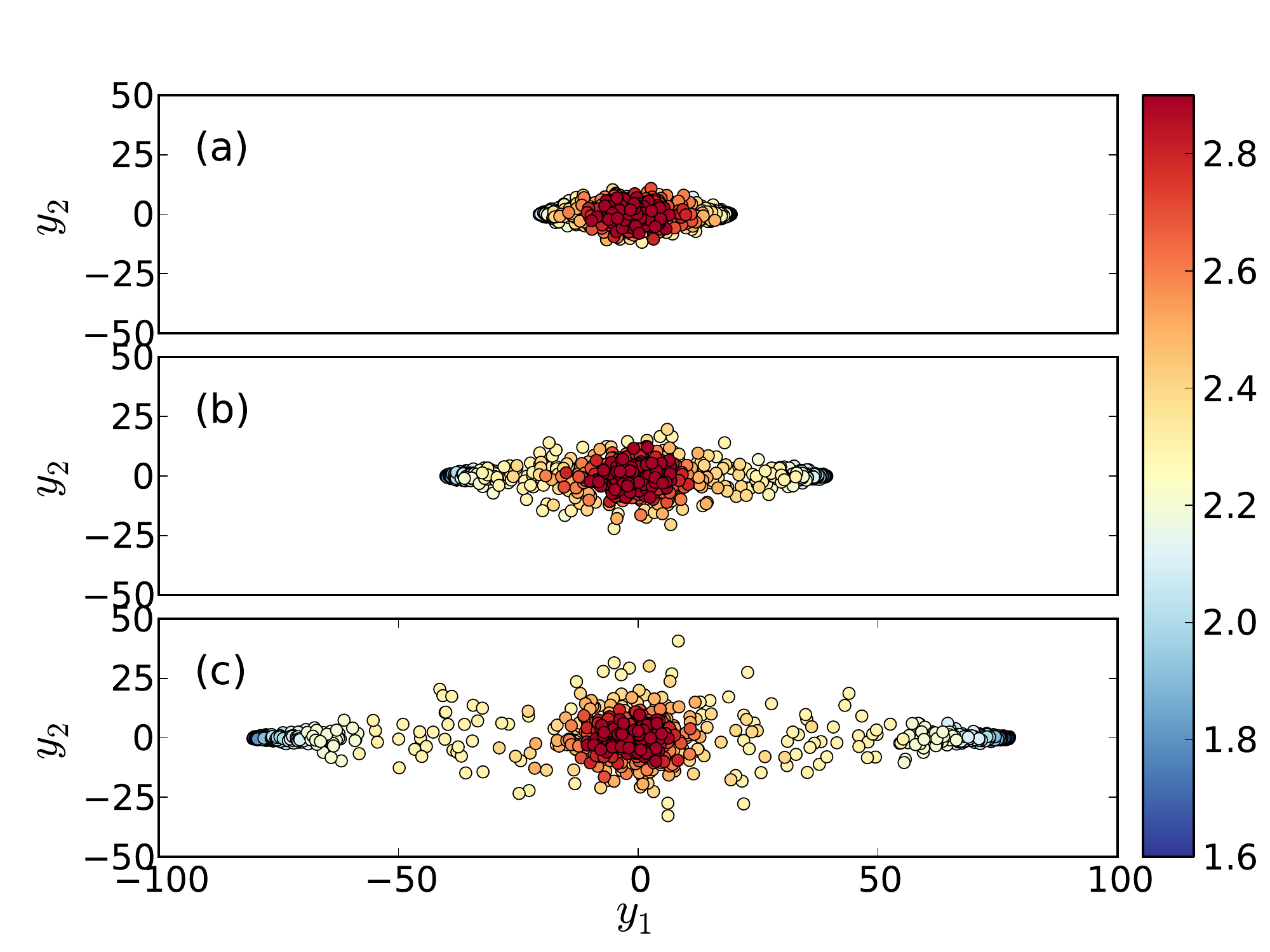}
  \caption{Projection of the samples onto the plane of the leading two principal components. The color bar on the right indicates the temperature $T/J$ of the samples. The panels (a-c) are for $N=20^{2},40^{2}$ and $80^{2}$ sites respectively.}
  \label{fig:projection}
\end{figure}

We write the orthogonal transformation into column vectors $W=(w_{1},w_{2},\ldots, w_{N})$ and denote $w_{\ell}$ as weights of the principal components in the configuration space. They are determined by an eigenproblem~\cite{Jolliffe:2002ku} \footnote{In practice this eigenproblem is often solved by singular value decomposition of $X$. In fact, replacing the input data $X$ (raw spin configurations collected at various temperature) by the wave function of a one-dimensional quantum system, the math here is identical to the truncation of Schmidt coefficients in the density-matrix renormalization group calculations~\cite{White:1992tg}. 
}
\begin{equation}
X^{T}X w_{\ell} = \lambda_{\ell}w_{\ell}.  
\label{eq:pca}
\end{equation}
The eigenvalues are nonnegative real numbers sorted in a descending order $\lambda_{1}\ge \lambda_{2}\ldots \ge \lambda_{N}\ge0$. Using the terminology of PCA, we denote the normalized eigenvalues $\tilde{\lambda}_{\ell}= \lambda_{\ell}/\sum_{\ell=1}^{N}\lambda_{\ell}$ as \emph{explained variance ratio}. When keeping only the first few principal components, PCA is an efficient dimension reduction approach which captures most variations of the original data. Moreover, PCA also yields an optimal approximation of the data in the sense of minimizing the squared reconstruction error~\cite{Jolliffe:2002ku}. 

Figure~\ref{fig:components} shows the first few explained variance ratios for various system sizes. Notably, there is only one dominant principal component. As the temperature changes the Ising configurations vary most significantly along the first principal component, whose weight is shown in the inset of Fig.~\ref{fig:components}. The flat distribution all over the lattice sites means the transformation actually gives the uniform magnetization $\frac{1}{N}\sum_{i} \sigma_{i}$. In this sense, PCA has identified the order parameter of the Ising model (\ref{eq:Ising}) upon a phase transition.

\begin{figure}[t]
  \centering
  \includegraphics[width=\columnwidth]{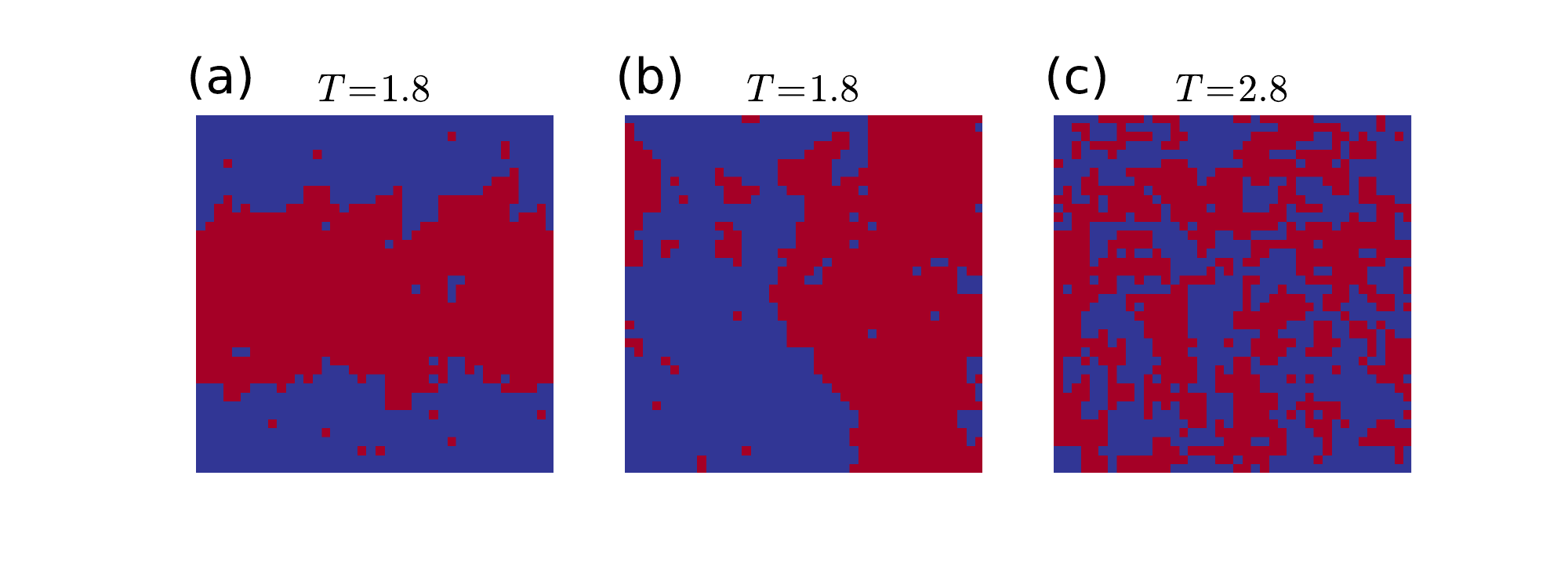}
  \caption{Typical configurations of the COP Ising model at below (a,b) and above (c) the critical temperature. Red and blue pixels indicate up and down spins. There are exactly half of the pixels are red/blue due to the constraint $\sum_{i}\sigma_{i}\equiv0$.}
  \label{fig:COPIsing_configs}
\end{figure}

Next, we project the samples in the space spanned by the first two principal components, shown in Figure~\ref{fig:projection}. The color of each sample indicates its temperature. The projected coordinates are given by the matrix-vector product
\begin{equation}
y_{\ell} = X w_{\ell}. 
\label{eq:projection}
\end{equation}
The variation of the data along the first principal axis $y_{1}$ is indeed much stronger than that along the second principal axis $y_{2}$. Most importantly, one clearly observes that as the system size enlarges the samples tend to split into three clusters. The high-temperature samples lie around the origin while the low-temperature samples lie symmetrically at finite $y_{1}$. The samples at the critical temperature (light yellow dots) have broad spread because of large critical fluctuations. We note that Ref.~\cite{Carrasquilla:2016wu} presents a different low dimension visualization of the Ising configurations using stochastic neighbor embedding technique. 

When folding the horizontal axis of Fig.~\ref{fig:projection} into $\sum_{i} |\sigma_{i}|$ or $(\sum_{i}{\sigma_{i}})^{2}$ the two clusters associated with the low-temperature phase merge together. With such a linear separable low dimensional representation of the original data, a cluster analysis~\footnote{See, for example, the methods provided in the scikit-learn cluster module \url{http://scikit-learn.org/stable/modules/clustering.html} } can easily divide the samples into two phases, thus identifying the phase transition. 
Notice that our unsupervised learning analysis does not only finds the phase transition and an estimate of the critical temperature but also provides insight into the order parameter. 

\begin{figure}[t]
  \centering
  \includegraphics[width=\columnwidth]{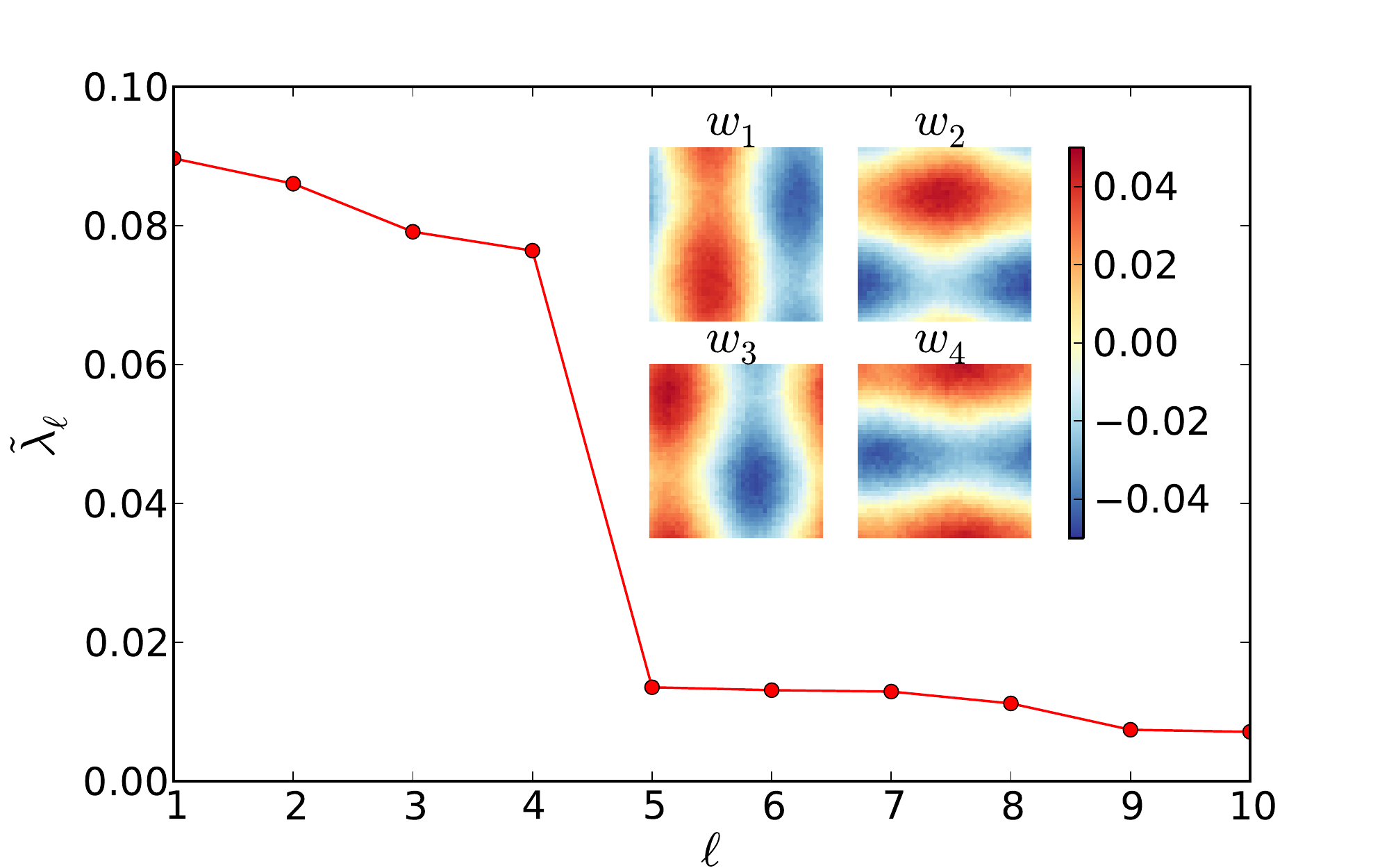}
  \caption{Explained variance ratios of the COP Ising model. Insets show the weights corresponding to the four leading principal components.}
  \label{fig:COPIsing_components}
\end{figure}

Having established the baseline of applying the unsupervised learning techniques in the prototypical Ising model, we now turn to a more challenging case where the learner can make nontrivial findings. For this, we consider the same Ising model \Eq{eq:Ising} with a conserved order parameter (COP) $\sum_{i}\sigma_{i}\equiv 0$. This model describes classical lattice gasses~\cite{Newman:1999fd}, where the occupation of each lattice site can be either one or zero and the particles interact via a short-range attraction. The conserved total magnetization corresponds to the constraint of a half filled lattice. 

On a square lattice with periodic boundary conditions, the spins tend to form two domains at low-temperatures shown in Fig.~\ref{fig:COPIsing_configs}(a,b). The two domain walls wrap around the lattice either horizontally or vertically to minimize the domain wall energy~\cite{Newman:1999fd}. Besides, the domains can also shift in space due to translational invariance. As the temperature increases, these domain walls melt and the system restores both the translational and rotational symmetries in the high-temperature phase shown in  Fig.~\ref{fig:COPIsing_configs}(c). At zero total magnetization, the critical temperature of such solid-gas phase transition is the same as the Ising transition $T_{c}/J\approx2.269$~\cite{Yang:1952fh}. However, since the total magnetization is conserved, simply summing up the Ising spins can not be used as an indicator to distinguish the two phases. In fact, it is unclear to the author which quantity signifies the phase transition before this study. It is, therefore, a good example to demonstrate the ability of the unsupervised learning approach. 

\begin{figure}[t]
  \centering
  \includegraphics[width=\columnwidth]{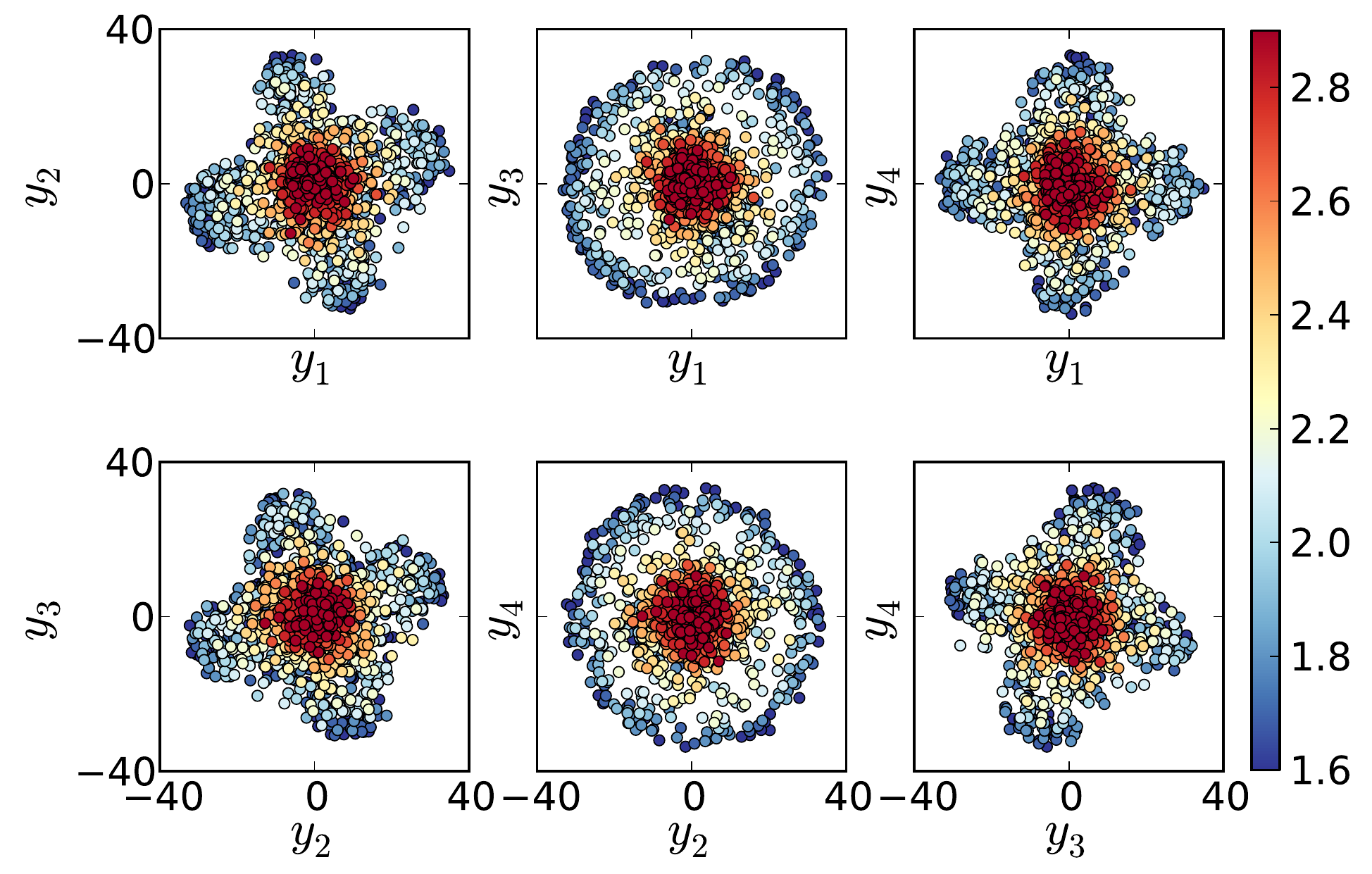}
  \caption{Projections of the COP Ising samples to the four leading principal components.} 
  \label{fig:COPIsing_projection}
\end{figure}

We perform the same PCA on the COP Ising configurations sampled with Monte Carlo simulation~\cite{Newman:1999fd} and show the first few explained variance ratios in Fig.~\ref{fig:COPIsing_components}. Notably, there are four instead of one leading principal components. Their weights plotted in the insets of Fig.~\ref{fig:COPIsing_components} show notable nonuniformity over the lattice sites. This indicates that in the COP Ising model the spatial distribution of the spins varies drastically as the temperature changes.  Denote Euclidean coordinate of site $i$ as $(\mu_{i},\nu_{i})$, where $\mu_{i}, \nu_{i}=1,2,\ldots,\sqrt{N}$. The weights of the four leading principal components can be written as $\cos(\theta_{i}), \cos(\phi_{i}), \sin(\theta_{i}),\sin(\phi_{i})$, where $(\theta_{i}, \phi_{i})= (\mu_{i}, \nu_{i})\times2\pi /\sqrt{N}$~\footnote{The weights shown in the inset of Fig.~\ref{fig:COPIsing_components} are linear mixtures of them.}.  Note these four mutually orthogonal weights correspond to the two orientations of the domain walls shown in Fig.~\ref{fig:COPIsing_configs}(a,b). Therefore, the PCA correctly finds out the rotational symmetry breaking caused by the domain wall formation. 

\begin{figure}[t]
  \centering
  \includegraphics[width=\columnwidth]{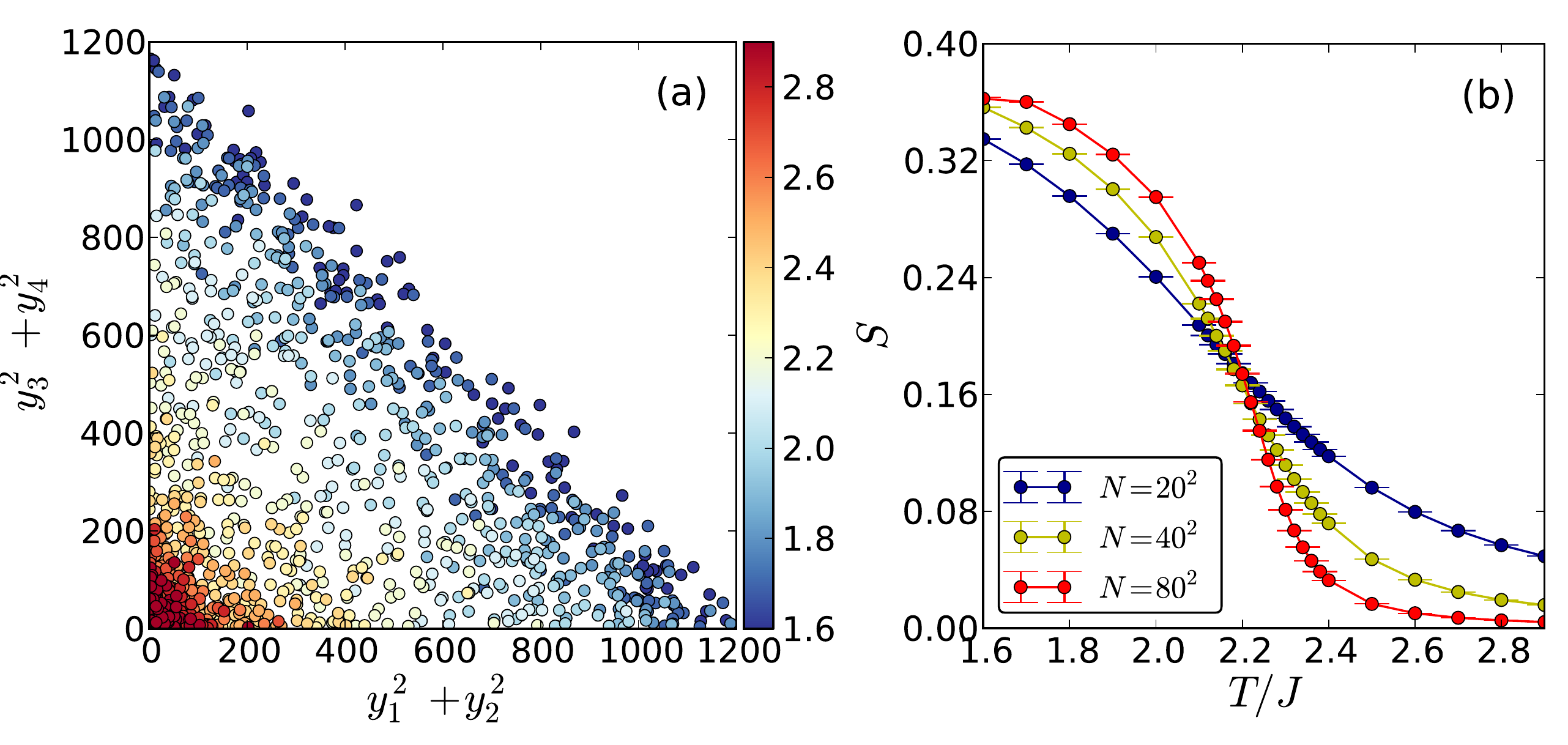}
  \caption{(a) Further projection of the COP Ising samples to a two-dimensional space. (b) The structure factor \Eq{eq:S} of the COP Ising model versus temperature for various system sizes.} 
  \label{fig:S}
\end{figure}

To visualize the samples in the four-dimensional feature space spanned by the first few principal components, we plot two-dimensional projections in Fig.~\ref{fig:COPIsing_projection}. In all cases, the high-temperature samples are around the origin while the low-temperature samples form a surrounding cloud. Motivated by the circular shapes of all these projections, we further reduce to a two-dimensional space via a nonlinear transformation $(y_{1},y_{2},y_{3},y_{4}) \mapsto (y_{1}^{2}+ y_{2}^{2}, y_{3}^{2}+ y_{4}^{2})$. As shown in Fig.~\ref{fig:S}(a), the line $\sum_{\ell=1}^{4} y_{\ell}^{2}=\mathrm{const}$ (a four dimensional sphere of a constant radius) separates the low and high temperature samples. This motivates a further dimension reduction to a single variable $\sum_{\ell=1}^{4} y_{\ell}^{2}$ as an indicator of the phase transition in the COP Ising model. 


Substituting weights of the four principal components $\cos(\theta_{i}), \cos(\phi_{i}), \sin(\theta_{i}),\sin(\phi_{i})$, the sum $\sum_{\ell=1}^{4}y_{\ell}^{2}$ is proportional to 
\begin{equation}
S =\frac{1}{N^{2}}\sum_{i,j}\sigma_{i}\sigma_{j} \left[\cos\left(\theta_{i}-\theta_{j}\right) + \cos\left(\phi_{i}-\phi_{j}\right) \right]. 
\label{eq:S}
\end{equation}
Even though such structure factor was unknown to the author before it was discovered by the learner, one can convince himself it indeed captures the domain wall formation at low temperatures shown in Fig.~\ref{fig:COPIsing_configs}(a,b). Figure~\ref{fig:S}(b) shows the structure factor versus temperature for various system sizes. It decreases as the temperature increases and clearly serves as a good indicator of the phase transition. We emphasis that the input spin configurations contain no information about the lattice geometry nor the Hamiltonian. However, the unsupervised learner has by itself extracted meaningful information related to the breaking of the orientational order. Therefore, even without the knowledge of the lattice and the analytical understanding of the structure factor \Eq{eq:S}, $\sum_{\ell=1}^{4}y_{\ell}^{2}$ plays the same role of separating the phases in the projected space. 




It is interesting to compare our analysis of phase transitions to standard imagine recognition applications. In the Ising model example, the learner essentially finds out the brightness of the imagine $\sum_{i}\sigma_{i}$ as an indicator of the phase transition. While in the COP Ising model example, instead of detecting sharpness of the edges (melting of domain walls) following the ordinary imagine recognition routine, the PCA learner finds out the structure factor \Eq{eq:S} related to symmetry breaking, which is a fundamental concept in phase transition and condensed matter physics.

Considering PCA is arguably one of the simplest unsupervised learning techniques, the obtained results are rather encouraging. In essence, our analysis finds out the dominant collective modes of the system related to the phase transition. The approach can be readily generalized to more complex cases such as models with emergent symmetry and order by disorder~\cite{Moessner:2001da}. The unsupervised learning approach is particularly profitable in the case of hidden or multiple intertwined orders, where it can help to single out various phases. 

Although \emph{nonlinear} transformation of the raw configuration \Eq{eq:S} was discovered via visualization in Fig.~\ref{fig:COPIsing_projection}, simple PCA is however limited to linear transformations. Therefore, it remains challenging to identify more subtle phase transitions related to the topological order, where the indicators of the phase transition are nontrivial nonlinear functions of the original configurations. For this purpose, it would be interesting to see if a machine learning approach can comprehend concepts such as duality transformation~\cite{Wegner:1971jf}, Wilson loop~\cite{Wilson:1974ji} and string order parameter~\cite{denNijs:1989dp}. A judicial apply of kernel techniques~\cite{Scholkopf:1998vb} or neural network based deep autoencoders~\cite{Hinton:2006kc} may achieve some of these goals. 


Furthermore, although our discussions focus on thermal phase transitions of the classical Ising model, the unsupervised learning approaches can also be used to analyze quantum many-body systems and quantum phase transitions~\cite{Sachdev:2011uj}. In these applications, diagnosing quantum states of matter without knowledge of Hamiltonian is a useful paradigm for cases with only access to wave-functions or experimental data. 




\paragraph{Acknowledgment}
The author thanks Xi Dai, Ye-Hua Liu, Yuan Wan, QuanSheng Wu and Ilia Zintchenko for discussions and encouragement. The author also thanks Zi Cai for discussions and careful readings of the manuscript. L.W. is supported by the start-up funding of IOP-CAS.  

\bibliographystyle{apsrev4-1}
\bibliography{IsingML}


\end{document}